\documentclass{PoS}
\usepackage{amssymb,amsmath}
\title{Indirect detection of Dark matter with gamma-rays - status and perspectives}
\newcommand{\sigmav}{<\sigma\,v>}
\newcommand{\cms}{cm^3s^{-1}}

\ShortTitle{Indirect detection of Dark matter with gamma-rays}

\author{\speaker{Jan Conrad}\\
        Oskar Klein Centre for Cosmoparticle Physics, Department of Physics, Stockholm University\\
        E-mail: \email{conrad@physics.su.se}}

\abstract{In my contribution I review the status of indirect detection of dark matter with gamma-rays, including results of the Fermi Gamma-ray Space Telescope as well as Imaging Air Cherenkov Telescopes (IACT), like H.E.S.S., MAGIC and VERITAS. I will also briefly address the  perspectives for the next generation IACT, the Cherenkov Telescope Array.}

\FullConference{Identification of Dark Matter 2010-IDM2010\\
		July 26-30, 2010\\
		Montpellier France}

\begin{document}

\section{Introduction}
Observations from Galactic to cosmological scales leave little room to explanations for Dark Matter (DM) other than that it is provided by a new type of particle. The most studied option for this particle is a Weakly Interacting Massive Particle (WIMP) which without much fine tuning provides a relic abundance which is in the right ball park to account for the inferred amount of DM, sometimes referred to as the "WIMP miracle"\footnote{we will in this contribution concentrate on WIMP DM and leave other candidates, in particular axions or axion-like particles, aside.}. WIMPs are searched for in a variety of ways: by particle production at the LHC, by searching for nuclear recoils signals and last but not least by searching for a signal in secondary products of WIMP annihilation or decay, in particular gamma-rays.\\

\noindent  
Spectral signatures in gamma-rays can be classified in smoking gun signals and ambiguous signals. The smoking gun signals are mainly strong spectral features, such as bumps from virtual internal brems-strahlung \cite{Bringmann:2007nk} or lines from loop processes, e.g. \cite{Ullio:2002pj}, ambigous signals are due to continuum emission from pion decay, e.g. \cite{Cesarini:2003nr} or hard power-laws (e.g. from final state radiation, e.g.  \cite{Birkedal:2005ep}). These signal types can still be useful if many sources are combined (see below).\\

\noindent
A signature which has studied in more detail recently \cite{Ando:2006cr}\cite{Cuoco:2007sh}\cite{SiegalGaskins:2008ge}\cite{Cuoco:2010jb} and has also been discussed in this conference \cite{SiegalGaskins:2010nh} is the spatial distribution of gamma-rays. This signature will not be further discussed here.

\section{What happened since IDM 2008?}

The most important event since last IDM (in 2008 in Stockholm), as far as gamma-ray detection of DM is concerned,  was the arrival of data from the Fermi Large Area Telescope (Fermi-LAT) \cite{Atwood:2009ez}, that began nominal operation in August 2011 (essentially during the last IDM). At IDM 2008, Fermi-LAT did promise exciting results to come, and the promises were kept. In 2009, the hype was a TeV WIMP, ignited by the rising position ratio of PAMELA \cite{Adriani:2008zr} which together with the hard electron+positron spectrum presented by Fermi \cite{Abdo:2009zk} indicated the presence of an additional positron source, which could be DM annihilation or decay (then most likely heavy WIMP ($\sim$ TeV), see e.g.\cite{Bergstrom:2009fa}). Pulsars on the other hand constitute a particular well motivated conventional explanation \cite{Malyshev:2009tw}. Since the PAMELA/Fermi-LAT results require annihilation (or decay) primarily into leptonic final states, attempts to explain the PAMELA result by WIMPs revived the consideration of another gamma-ray signature: inverse Compton of the produced leptons on IR/CMB background light  \cite{Baltz:2004bb} \cite{Colafrancesco:2005ji}. This signature is fundamentally different from those I presented above, since it depends on the astrophysical environment of the DM source, it can be dominant for models with preferentially leptonic decays, see e.g. \cite{Profumo:2009uf}.\\

\noindent
As eager the community was to explain data with high mass WIMPs in 2009, as popular became low mass WIMPs {$\sim$ 10 GeV} in 2010,  mainly due to results by CoGenT \cite{Aalseth:2010vx} and the already famous DAMA/Libra \cite{Bernabei:2008yi} detection. Subsequently low mass WIMPs could also be "seen" in gamma-rays (e.g. \cite{Hooper:2010mq}). It is likely that all of these indications will turn out to be astrophysical or instrumental backgrounds, and that we did not come closer to finding the "WIMP". However, it is certainly preferable to have accurate data to get confused by, than to be left to mere speculation. And this will become worse (or better, depending on how you see it) with the new data from the Large Hadron Collider.

\section{Present and future gamma-ray experiments.}

\begin{figure}
	\centering
		\includegraphics[width=0.75\textwidth]{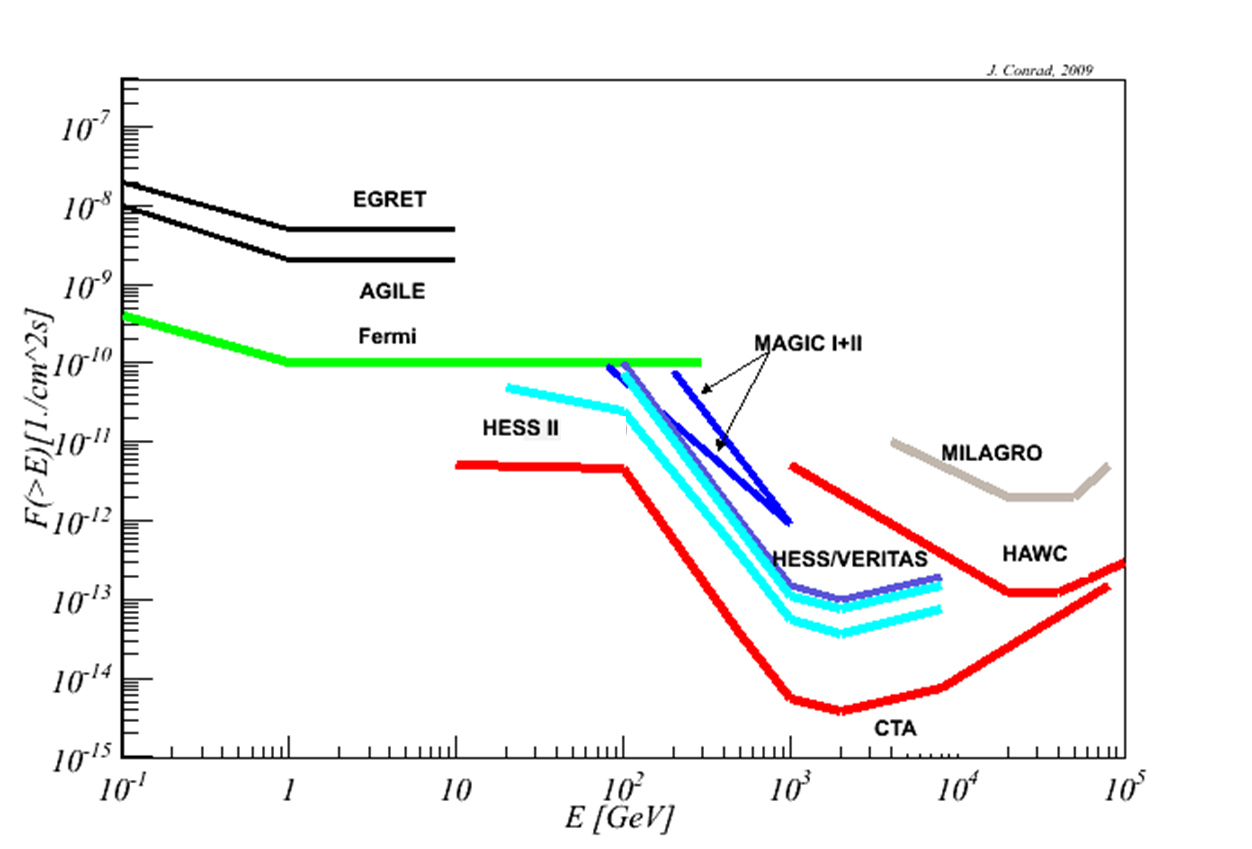}
\caption{Integral flux sensitivity estimates as function of gamma-ray energy for existing and planned gamma-ray telescopes. MAGIC II is already operating, the estimated time for operation for H.E.S.S. II is 2012, for CTA 2018. Note that sensitivity curves shown here might be somewhat 
outdated, but reflect approximately the relative sensitivities.}
\label{fig:Bild5}
\end{figure}

Figure \ref{fig:Bild5} shows the integral flux sensitivity of experiments for astrophysical gamma-ray detection as
a function of gamma-ray energy. The satellite experiments Fermi and the former EGRET, pair conversion telescopes with
a tracker and a calorimeter, cover the energy range below roughly 300 GeV. In the energy range above roughly 100 GeV, the detection area for pair conversion telescopes (order 1 $m^2$) becomes too small and Imaging Air Cherenkov Telescopes (IACTs) are competetive. The most relevant ones are H.E.S.S. \cite{HESS}, VERITAS \cite{VERITAS} and MAGIC \cite{MAGIC}. H.E.S.S. is in the process of upgrading the four telescope configuration with a fifth telescope large mirror ($28 m$ diameter) telescope, MAGIC-II (adding an additional telescope) is operating and initial results show that it performs to expectations. For comparison the predicted sensitivity for the next generation experiment CTA (Cherenkov Telescope Array) is shown \footnote{The recent design report \cite{Consortium:2010bc}  shows that the sensitivity below 50 GeV is probably a factor 3-5 smaller than in this figure.}. The energy range above 10 TeV is covered mainly by water Cherenkov telescopes such as MILAGRO \cite{MILAGRO} and the future experiment HAWC \cite{HAWC}.\\

\noindent
Except for detection area, important performance features are the energy resolution and (to lesser extent for DM searches) the pointing resolution. Fermi-LAT and IACTs have rather similar energy resolution of between $\sim$ 8 \% and $\sim 10 \%$.\\

\noindent
The satellite experiments are wide field of view (FOV) experiments, Fermi-LAT having a field of view of 2.4 sr. Ground based Cherenkov telescopes on the other hand have a small FOV ($\sim  10^{-2}$ sr) and therefore need to be pointed at promising targets. This is an inherent problem for DM searches since observation time is limited and one has to compete with more "conventional" astrophysical sources, which are guaranteed gamma-ray emitters. Typically only 5-8 \%  of the total available observation time per year is dedicated to DM searches per se. As a recent study indicates a dedicated DM gamma-ray detector might be argued for \cite{Bergstrom}.

\section{Targets}
The number of photons created by DM particle annihilation is proportional to the square of the DM density along the line of sight\footnote{for decaying dark matter the dependence is linear.}. This motivates a number of promising targets for DM searches, namely those with known density enhancements, foremost the Galactic Centre and close-by dwarf galaxies and galaxy clusters. In addition, integrating over density enhancements -- Galactic or extragalactic -- may yield a diffuse contribution.

\subsection{Galactic Centre}
The Galactic Centre (GC) is expected to be one of the strongest sources for gamma-ray radiation due to DM annihilation. At the same time, the GC is crowded with conventional gamma-ray sources: HESS and Fermi-LAT sources at the GC are consistent with each other and Sgr A* and PWN 359.9 \cite{Acero:2010}\cite{Cohen-Tanugi:2009}. H.E.S.S. has used the spectrum of the GC sources (which is inconsistent with being dominantly due to DM), to constrain DM models \cite{Ripken:2007}\cite{Ripken:2011}, which allows constraining models with extreme properties (for example those having a large contribution from virtual internal bremsstrahlung) and under assumptions of very cuspy profiles. An updated H.E.S.S. analysis approach has been presented in this conference \cite{Nekrassov:2010} and results subsequently in \cite{Abramowski:2011hc}. Here the GC observations have been used and the source region was $1^\circ$ centered on the GC , but the galactic plane is excluded (a variant of the halo analysis described below, the difference being that the background estimate is obtained from within the field of view). Constraints obtained are about an order of magnitude better than those presented for the Sagittarius dwarf (see below).\\

\noindent
The Fermi-LAT has still not presented any constraints from the GC in a refereed publication. Spectral results presented at conferences, e.g. \cite{Vitale:2009} do present a relatively weak excess between 3 and 5 GeV with respect to the most simple models of the local sources and Galactic diffuse emission, which lead to claims that this signal is consistent with DM \cite{Hooper:2010mq} but also more sophisticated models for the sources and the diffuse emission have been proposed for its explanation, see e.g. \cite{Boyarsky:2010dr}.

\subsection{Dwarfs, Dark Matter satellites and Galaxy Clusters}
\begin{figure}
	\centering
		\includegraphics[width=0.75\textwidth]{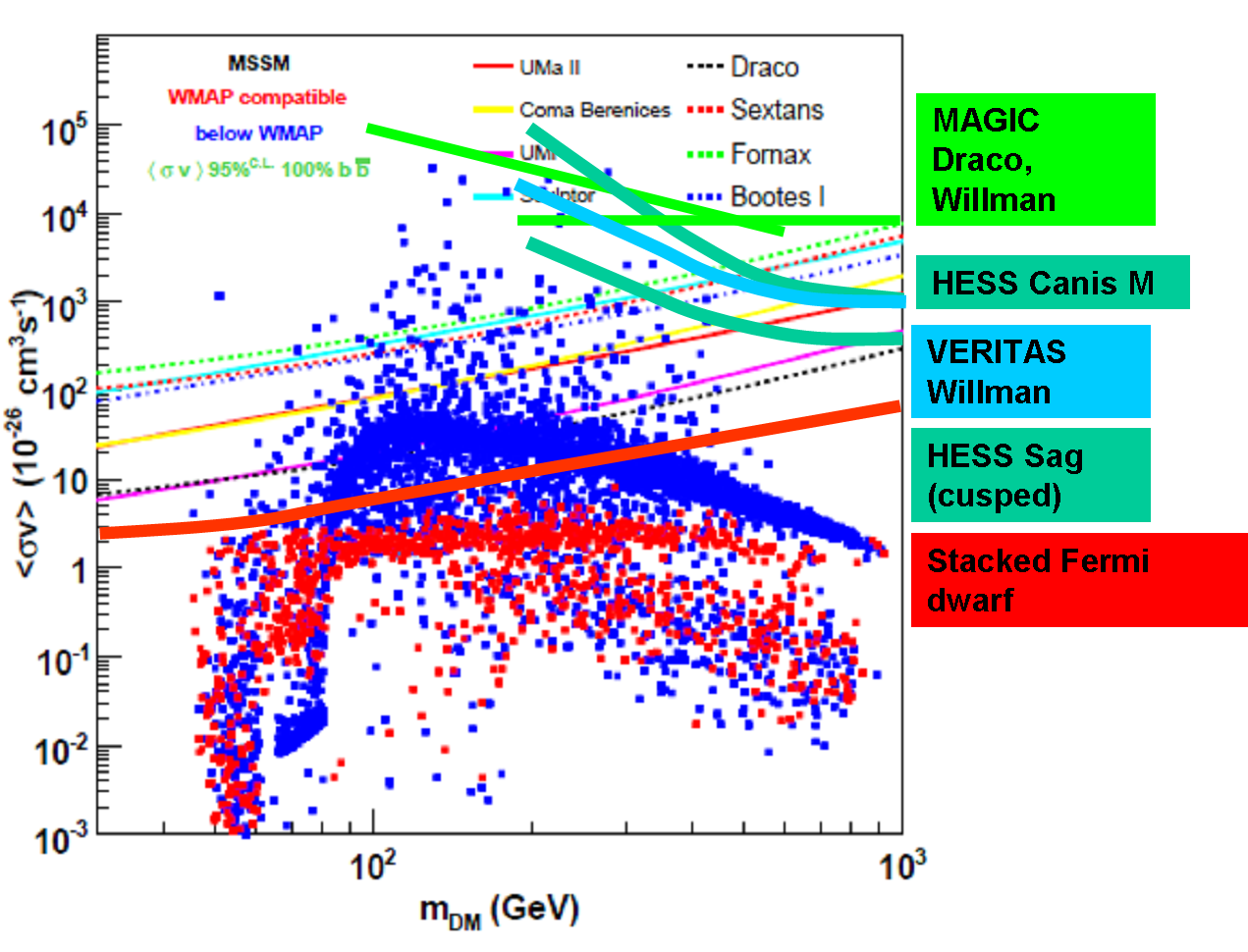}
	\caption{Velocity averaged annihilation cross-sections as function of the mass of the WIMP with current constraints from observations of dwarf galaxies. The figure is adopted from \cite{Abdo:2010ex} with the dwarf constraints from the ground based experiments and the stacked Fermi dwarf overlaid. The single dwarf constraints from the Fermi-LAT are derived assuming 100 \% annihilation into $b\bar{b}$, for the others, see the references given in the main text. None of these limits includes uncertainties on the DM density. For the stacked dwarf, the final results are envisaged to include those. The figure is not doing full justice to IACT as their constraints are expected to be better than those of the Fermi-LAT above 1 TeV, constraints are thus complementary in terms of expected WIMP mass.}
\label{fig:status}	
\end{figure}

About 20 Dwarf spheroidal Galaxies, satellites of our own Galaxy have been targeted by the main gamma-ray experiments. Fermi-LAT presented observations of 14 Dwarf galaxies with DM constraints for about 2/3 of them based on availability of reasonable estimates of the DM distributioin based on stellar data (see figure \ref{fig:status}) \cite{Abdo:2010ex}. H.E.S.S presented constraints from the Sagittarius Dwarf \cite{Aharonian:2007km} and Canis Major \cite{Aharonian:2009a} \footnote{more recently also Sculptor and Carina \cite{Acero:2010zzt}. }, MAGIC for Willman I and Draco \cite{Albert:2007xg} \cite{Aliu:2008ny} \footnote{more recently also Segue I \cite{Aleksic:2011jx}.}, Veritas for Bootes I, Draco, Ursa Minor and Willman I \cite{veritas:2010pja}.  The typical exposure of the dwarfs for the ground-based experiments is at about 10-20 hours, which can be compared to an equivalent 1000 hours (11 month of survey mode) of Fermi, which therefore provides comparable constraints even at 1 TeV assumed particle mass. However,
 it should be emphasized that in the region above 1 TeV the IACT will be more sensitive than the Fermi-LAT, thus truely complementary to Fermi-LAT.  Constraints derived from single dwarf analyses are about one order of magnitude larger than the interesting cosmological bench mark $\sigmav \sim 3 \times 10^{26} cm^{3}s^{-1}$. It is also noteworthy, that it has recently been claimed that the H.E.S.S. constraint might be over-optimistic by almost one order of magnitude \cite{Viana:2011tq}, due to updated and more accurate estimates of its DM density.\\

\noindent
A particular interesting analysis approach is the stacking of dwarf galaxies, which due to the universality of the DM spectrum in dwarf galaxies, can be performed as a combined likelihood ("likelihood stacking") analysis \cite{Garde:2011wr}. Not only is the statistical power increased, but also the impact of individual uncertainties in dark matter density (which enters the expected flux quadratically), is reduced. It can therefore be argued that this approach presents at the moment the strictest, yet most robust, limits on DM induced gamma-ray signal. Preliminary results have been presented at this conference, for the first time approaching the canonical cosmological cross-section. This looks exciting, but does not include statistically the uncertainties in the dark matter density -- the same is true for virtually all DM searches, so to see this implemented is a step forward regardless of which analysis it is applied to. Fermi-LAT has promised to present constraints including these uncertainties soon.\\

\noindent
Finally, there could be a  class of objects known as "Dark Matter satellites", Galactic satellites void of gas and stars, without a counter-part in other wavelength are searched for, either in blind searches \cite{Wang:2009} or concentrating on known, unassociated sources reported by the Fermi-LAT \cite{Buckley:2010vg}. Such sources with interesting properties could then be followed up with ground based telescopes. No detection has been reported.\\

\noindent
Galaxy clusters are dark matter dominated and much more massive than dwarf galaxies. Being situated at high-lattitude analysis-wise they could be considered as similar to dwarf spheroidals, since they seem to have similar backgrounds. However, galaxy clusters are very likely sources of cosmic ray induced gamma-rays (which, however, have not been detected, yet), thus exhibit an in-situ background to be taken into account\footnote{another difference to analysis of dwarf spheroidals is, that you might want to model galaxy clusters as extended sources.}.
Constraints reported by the Fermi-LAT \cite{Ackermann:2010rg} are in general weaker than for dwarf spheroidals. MAGIC presented results on the Perseus cluster \cite{Aleksic:2009ir}, which are 4 orders of magnitude from supersymmetric scenarios which are consistent with relic density measurements i.e. the constraint seems roughly one order of magnitude worse than the worst constraint presented by Fermi-LAT.

\subsection{Galactic halo emission}
The Galactic halo is a promising target due the vast number of photons produced and due to the fact that both the spectral and spatial distribution can be used as a discriminant. On the other hand, the conventional astrophysical background is particularly complicated. In order to obtain meaningful constraints, theoretical uncertainties in this background need to be quantitatively addressed. A task which is very challenging: consider for example, that in the times of the existence of the EGRET excess, Galactic diffuse emission was able to explain the excess or not depending basically on the assumption of the local CR spectrum being representative of the Galaxy or not ("conventional" vs "optimized" galprop model, see  \cite{Strong:1998fr}\cite{Strong:2004de}).  It is obvious that theoretical freedom of this sort is hard to cast into a quantitative statement.
Fermi-LAT has presented two approaches at this conference \cite{Zaharijas:2010ca}\cite{Anderson:2010hh}.\\

\noindent
One approach \cite{Anderson:2010hh} is to map the likelihood of at least part of the free parameters (diffusion coefficient, halo height, index of the energy dependence of the diffusion coefficient, assumed source distribution etc, etc) using CR data at the same time as the gamma-ray data. From the resulting likelihood function relatively honest constraints on a possible DM contribution could be obtained (the above described freedom of dismissing the local CR spectrum as representative is however not included in this approach). This approach is technically challenging and in addition structures observed by the Fermi-LAT (the Fermi-LAT lobes, for example), complicate the picture. But it is worthwhile: final results of this analysis would probably be the most accurate constraints obtainable from the Galactic halo with the Fermi-LAT (though they probably will not be the most efficient in constraining particle properties of DM).\\

\noindent
The other approach -- somewhat less ambitious-- \cite{Zaharijas:2010ca} uses a multi-component fit with 7 free parameters (normalizations of H2/H1 maps, Dark matter and normalization of the IC component and residual contribution from point sources), other diffusion parameters (e.g. halo height) are investigated with respect to their impact on the limits and the model allowing the largest DM contribution (required to be consistent with cosmic ray measurements) is chosen as a background model. Both analyses are work in progress, but are worth mentioning, since they represent important progress from more simplistic approaches (as for example just assuming a best-fit Galactic diffuse model without considering any uncertainties). \\

\noindent
H.E.S.S. has presented an on-source/off-source approach placing observations close to the GC (e.g. 2 deg) off-set and subtracting those farther away (e.g. 8 deg). Initial studies (assuming a density distribution according to the Aquarius simulation (e.g. \cite{Springel:2008zz}) indicate a sensitivity, which seems more promising than for example dwarf searches, e.g. \cite{Nekrassov:2010}. If systematics can be controlled, this might be the most promising way (so far) for ground based experiments to constrain DM \footnote{However, what a source stacking approach could accomplish for IACT analyses still remains to be seen.}.

\subsection{Isotropic gamma-ray emission}
Fermi-LAT has reported a measurement of the isotropic gamma-ray emission consistent with a single rather soft power law in energies from 40 MeV to 100 GeV \cite{Abdo:2010nz}. DM annihilation integrated over all halos could contribute  to the gamma-ray emission and would be quasi-isotropic, see e.g.\cite{Ullio:2002pj}. This measurement can thus be used to constrain DM particle properties. However, in order to do this, one needs not only to model DM substructure but also its cosmological evolution as well as intergalactic absorption - subject of research in itself. Resulting constraints are therefore highly model dependent - the virtue of this search being the rather low -- and temporally decreasing- background. Fermi-LAT has presented constraints on DM \cite{Abdo:2010dk} based on the data points in \cite{Abdo:2010nz}. Though this analysis is rather thorough, a consistent treatment of Galactic and extragalactic annihilation contribution, preferably also including a consideration of the predicted, conventional extragalactic background from --for example-- blazars, is still something to look forward too.

\subsection{Searches for spectral features}

Given the challenging backgrounds for DM searches, the possibility of detecting unambiguous spectral features is obviously attractive - some might find it the only way to be convinced beyond reasonable doubt. In vanilla DM models lines are suppressed due to the very nature of the DM particle, but there are models that 
Figure shows a summary of existing line limits from EGRET \cite{Pullen:2006sy}, Fermi \cite{Abdo:2010nc}\cite{Ylinen:2010} and H.E.S.S. \cite{Ripken:2007}. 


Current limits (see figure \ref{fig:lines}) are three orders of magnitude above the most generic predicted cross-sections predicted for line emission, though specific models are starting to be constrained (e.g.\cite{Mambrini:2010yp} \cite{Jackson:2009kg}).


\section{Direct constraints on Beyond Standard model physics and global fits}

A relatively novel approach which has been applied to both Fermi-LAT and H.E.S.S. data is to compare measured gamma-ray spectra directly with those predicted by one particular point in parameter-space of a given candidate of dark matter \cite{Scott:2009jn}\cite{Ripken:2011}. The theoretical model considered is hereby most often Minimal Supersymmetry in one of its simplest versions, i.e. 5 dimensional Constrained Minimal Supersymmetry (CMSSM), and inference on the parameter space of CMSSM is performed from a map of the full likelihood also including other measurements sensitive to supersymmetric contributions such as relic density, b->s$\gamma$, anomalous magnetic moment of the muon and various accelerator measurements as well as standard model parameters which are nuisance parameters in the prediction of SUSY observables (e.g. the top mass). These (as well as the CMSSM parameters not of interest in a 2 dimensional subspace representation of the full parameter space) are taken into account by a profile likelihood or likelihood marginalisation approach.\\

\noindent
Even for CMSSM (which is a relatively easy problem due to low dimensionality) the physics conclusions that can be drawn are at this point of limited interest, as is illustrated by a strong prior dependence of the present results. Nevertheless, given the high-dimensionality of the model space it is obvious that a combined analysis from different experiments with realistic experimental likelihoods ("a global fit"), including uncertainties on the background (whose model-space dimensionality can also be large, see above) and other nuisance parameters is desirable and it is good to see that tools for this task are rapidly developing.

\section{Next exit: the future}
Predicting the future is difficult, and inevitable backfire on those who do it. Nevertheless, let us make an attempt here. Fermi-LAT with some work and the sqrt(T) improvement in the upper limits may reach down to annihilation cross-sections of about a few times $\sigmav \sim 10^{-27} \cms$ at about 50 GeV,  the analysis of dwarf galaxies arguably being the most promising channel, with analysis of the Galactic halo as a contender (if there is any chance of seriously controlling the diffuse emission background). This is assuming 10 years of Fermi-LAT data and "some work" implies mainly increasing the effective area (already in development in the Fermi-LAT) and adding additional sources to the stacking (but no further, new analysis improvements).\\

\noindent
For large mass WIMPs (> 500 GeV -- 1 TeV ) ground based analysis of the halo seems to be the most promising venue, possibly reaching down to $\sigmav \sim  10^{-25} \cms $ with 50 hours of on-source/off-source observation. In the case of H.E.S.S., we are eagerly awaiting the first results from the 5 telescope configuration (H.E.S.S. II) which should be expected for 2012, which promises reduced energy threshold.\\

\begin{figure}
	\centering
		\includegraphics[width=0.75\textwidth]{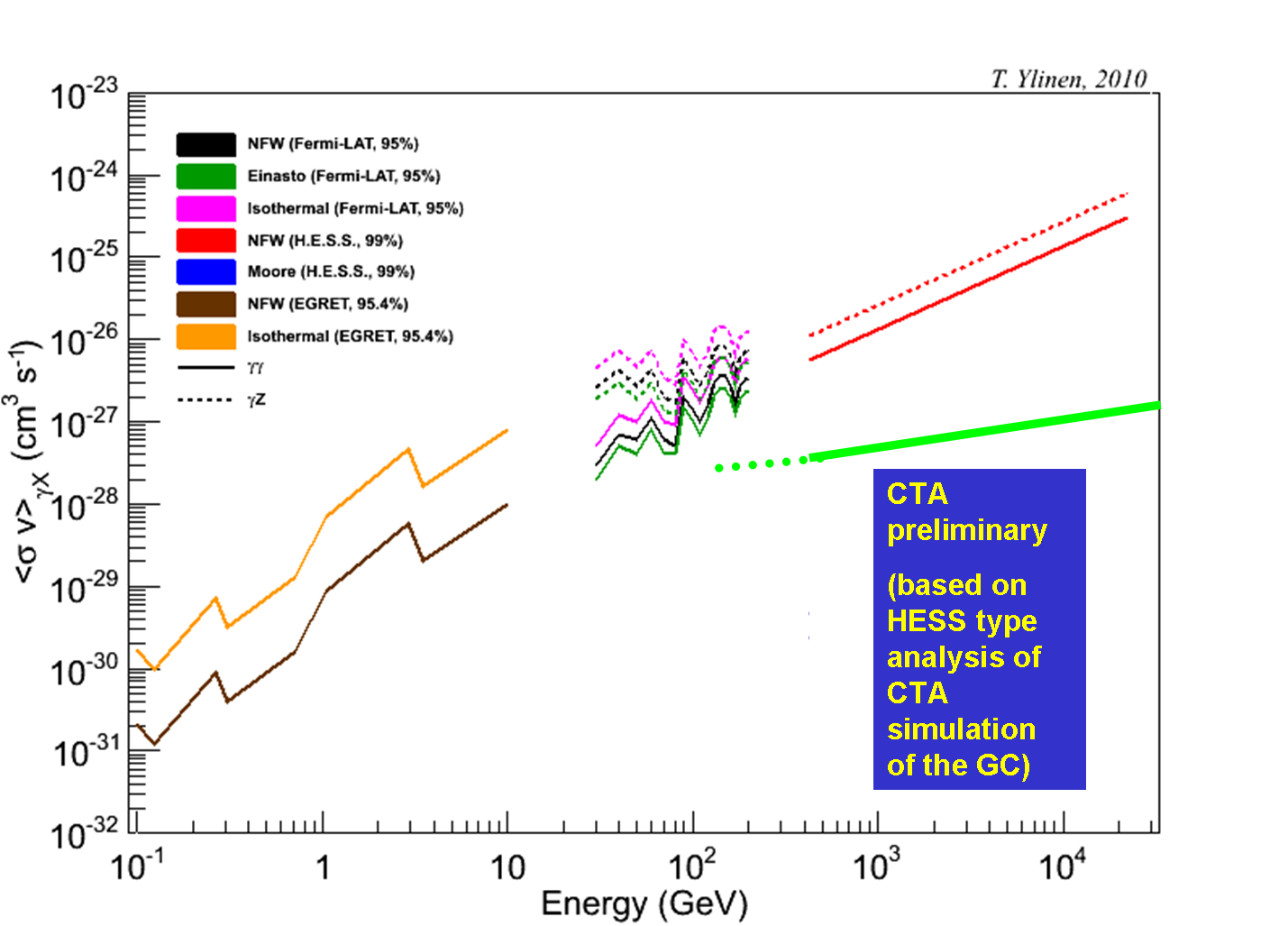}
	\caption{Velocity averaged annihilation cross-section into line emission and  EGRET, Fermi-LAT and H.E.S.S. constraints for different assumptions on density profile and final state. Also shown is a prediction for CTA. References in the text.}
\label{fig:lines}
\end{figure}

\noindent
The more distance future (post-Fermi) is dominated by the international effort to build the next generation air Cherenkov array (Cherenkov Telescope Array, CTA). Quantitative estimates, based on realistic array simulation are in production. Awaiting those, let us try to make some educated guesses. Recent estimates for the Sagittarius dwarf \cite{Viana:2011tq} lead to the conclusion that assuming the improvement in sensitivity for a halo analysis to scale as the improvement for Sagittarius dwarf would allow sensitivities to below $\sigmav \sim 10^{-27} \cms$ at masses as small as about 100 GeV, assuming boost in substructure as claimed by Aquarius and a total halo observation time of about 100 hours (50 on-source, 50 off-source). For the search for spectral lines, initial estimates of the sensitivity based on a H.E.S.S. type analysis of the GC, only using realistic CTA simulations, are shown in figure \ref{fig:lines} \footnote{the US part of the CTA consortium proposes an augmentation for the array which is not included in these studies and which might yield increased sensitivity especially for DM searches.}.
It is however noteworthy that CTA --in any case-- could  be able to cover parameter space  which is complementary (i.e. hard to access by direct detection experiments)  and indeed the point has been made that some part of the model space will be accessible only by a gamma-ray experiment (though then it will have to have sensitivities beyond CTA \cite{Bergstrom}). This is could in particular become true, should the LHC (by not detecting any DM candidates) push the WIMP mass to higher and higher values.


\section{Concluding remarks}
In this contribution I tried to give a brief summary of the status of searches for DM in gamma-rays, with emphasis on the developments between 2008 and 2010. It is obvious that the large amount of high quality data leads to increased number of claims for indications for dark matter, many of which are already ruled out before they are accepted publications. For DM detection in gamma-rays, the most important event since last IDM was the start of operation of the Fermi-LAT. In the mass range around 100 GeV, the constraints provided slowly start to approach the "miraculous" WIMP cross-section., i.e. close to $\sigmav \sim 3 \times 10^{26} cm^{3}s^{-1}$.  Novel approaches presented by ground based telescopes, that at the same time push their energy threshold to lower energies, may yield interesting limits. In all likelihood, the next 5 to 10 years will be very interesting from the point of view of detection of particle DM with gamma-rays.  Under reasonable assumptions on the astrophysical factors, we will probably be able to detect DM in gamma-rays at least in the most simple set-ups, i.e. those where a dominant fraction of DM is made of WIMPs which live within an order of magnitude of the canonical cross-section. This statement is arguably true only for the targets where the astrophysical backgrounds are controllable, such as dwarf galaxies. For other searches, even if likely to yield more gamma-rays from DM annihilation, such as the Galactic halo, but even the GC, the detection claim will be more difficult to make.\\

\noindent
Let me end with a disclaimer: the nature of this contribution is such that no complete account of the development can be presented. Choices have to be made and they are necessarily my own.

\section{Acknowledgements}
The author would like to thank Adrian Biland, Michele Doro, Agnieszka Jacholkowski and Luca Latronico for valuable comments on the manuscript. The author is Royal Swedish Academy Sciences Research fellow supported by a grant from the Knut and Alice Wallenberg foundation.

\end{document}